# PERFORMANCE IMPROVEMENT OF SPATIAL SEMANTIC SEGMENTATION WITH ENRICHED AUDIO FEATURES AND AGENT-BASED ERROR CORRECTION FOR DCASE 2025 CHALLENGE TASK 4

Technical Report

*Jongyeon Park[1], Joonhee Lee[2], Do-Hyeon Lim[1], Hong Kook Kim[1,2,*]*

*Hyeongcheol Geum[3], Jeong Eun Lim[3]*

[1] Dept. of AI Convergence, [2] Dept. of EECS,
Gwangju Institute of Science and Technology
Gwangju 61005, Korea
{{jypark3737, ljh13099, do-hyeon}@gm.,
hongkook@}gist.ac.kr

[3] AI Lab., R&D Center
Hanwha Vision
Seongnam-si, Gyeonggi-do 13488, Korea
{hch.geum, je04.lim}@hanwha.com

## ABSTRACT

This technical report presents submission systems for Task 4 of the DCASE 2025 Challenge. This model incorporates additional audio features (spectral roll-off and chroma features) into the embedding feature extracted from the mel-spectral feature to improve the classification capabilities of an audio-tagging model in the spatial semantic segmentation of sound scenes (S5) system. This approach is motivated by the fact that mixed audio often contains subtle cues that are difficult to capture with mel-spectrograms alone. Thus, these additional features offer alternative perspectives for the model. Second, an agent-based label correction system is applied to the outputs processed by the S5 system. This system reduces false positives, improving the final class-aware signal-to-distortion ratio improvement (CA-SDRi) metric. Finally, we refine the training dataset to enhance the classification accuracy of low-performing classes by removing irrelevant samples and incorporating external data. That is, audio mixtures are generated from a limited number of data points; thus, even a small number of out-of-class data points could degrade model performance. The experiments demonstrate that the submitted systems employing these approaches relatively improve CA-SDRi by up to 14.7% compared to the baseline of DCASE 2025 Challenge Task 4.

## 1. INTRODUCTION

The objective of spatial semantic segmentation of sound scenes (S5) is to jointly detect and separate multiple sound events from multichannel mixes [1][2]. This complex task requires a system to identify active sound classes (audio tagging) and to isolate their corresponding anechoic source signals accurately. The dual nature of this task presents significant challenges because optimal performance demands a model that can interpret the coarse semantic content for tagging and the fine-grained spectro-temporal details for separation.

A typical model often relies on a single feature representation, such as the mel-spectrogram, which can struggle to capture the diverse acoustic cues required for robust performance across all event classes. The performance of such a system highly depends on the training data quality. However, the challenge dataset is synthesized from a finite dataset of source recordings [3]. Hence, the model is apt to be sensitive to out-of-class or perceptually ambiguous samples, which can degrade the overall accuracy.

This work presents several systems submitted for Task 4 of the DCASE 2025 Challenge, incorporating three critical enhancements to address the challenges mentioned above. First, additional audio features, including spectral roll-off [4] and chroma [5], are incorporated into the embedding feature extracted from the mel-spectral feature to provide alternative perspectives on the audio that complement the mel-spectrogram. Second, to improve the final class-aware signal-to-distortion ratio improvement (CA-SDRi) [1][2] metric by identifying and rectifying false-positive (FP) predictions, an agent-based label correction system is proposed as a post-processor applied to the outputs of the tagging and separation modules. Finally, a meticulous dataset refinement strategy is performed to remove problematic samples and expand low-resource classes with external data by auditing the provided source data.

Following this introduction, Section 2 describes the dataset and data refinement process. Next, Section 3 details the proposed system architecture and agent-based correction system. Then, Section 4 presents the experimental setup and results. Finally, Section 5 concludes this report.

## 2. DATASET

This work for the DCASE 2025 Challenge Task 4 is based on the provided development dataset, synthesized from various sources using the SpatialScaper toolkit [6]. This section outlines the composition of the official dataset and the specific data curation and augmentation steps to improve the performance of the model.

### 2.1. Official dataset composition

[*] This work was supported in part by Hanhwa Vision Co. Ltd., by the GIST-MIT Research Collaboration grant, and by the IITP grant funded by the Korea government (MSIT; No.2022-0-00963).

Table 1: Summarization of the distribution of the training samples after refinement and augmentation

| Class | Original count | Duration ≥ 1.5 | Heterogeneous data | Added count | Final count |
|---|---|---|---|---|---|
| Alarm clock | 102 | 2 | 37 | - | 63 |
| Bicycle bell | 230 | 10 | 22 | - | 198 |
| Blender | 141 | 0 | 2 | - | 139 |
| Buzzer | 181 | 0 | 0 | - | 181 |
| Clapping | 482 | 195 | 67 | - | 220 |
| Cough | 443 | 0 | 8 | - | 435 |
| Cupboard open/close | 413 | 32 | 25 | - | 356 |
| Dishes | 399 | 99 | 61 | - | 239 |
| Doorbell | 75 | 4 | 24 | 51 | 98 |
| Footsteps | 388 | 53 | 16 | - | 319 |
| Hair dryer | 25 | 0 | 2 | - | 21 |
| Mechanical fans | 126 | 0 | 0 | - | 126 |
| Musical keyboard | 503 | 41 | 35 | 10 | 437 |
| Percussion | 2063 | 858 | 213 | - | 992 |
| Pour | 93 | 1 | 3 | - | 89 |
| Speech | 1211 | 511 | 68 | - | 632 |
| Typing | 436 | 28 | 8 | - | 400 |
| Vacuum cleaner | 66 | 0 | 0 | - | 66 |

This challenge provided distinct datasets for synthesizing the final audio mixes:
- Sound sources: The primary source for the 18 event classes was the Anechoic Sound Event 1K dataset, supplemented with sounds from Freesound Dataset 50k (FSD50K) [7] and Expressive Anechoic Recordings of Speech (EARS) [8].
- Room impulse responses: Spatial characteristics were introduced using the provided room impulse response datasets, including new recordings from NTT and the First-Order Ambisonics Room Impulse Response (FOA-MEIR) dataset [9].
- Background interference: Various noise sources were designated to create realistic soundscapes, including audio from the FOA-MEIR, FSD50K, ESC-50 [10], and DISCO [11] datasets.

All audio mixes of 10 seconds long each were generated at a sampling rate of 32 kHz. The number of active sound events in each mix ranged from one to three, with the signal-to-noise ratio of each event varying between 5 and 20 dB.

### 2.2. Data refinement and augmentation

During development, some samples in the sound source dataset hindered training performance; thus, a two-stage data refinement strategy was implemented. First, based on the observation that event durations shorter than 1.5 s adversely degraded classification performance, these short samples were excluded from the training set. Second, all sound event samples were manually examined, removing perceptually heterogeneous samples that did not align with the acoustic characteristics of their respective classes.

Although this refining process was generally beneficial, it led to a notable performance degradation for some specific classes, such as the 'Doorbell' and 'MusicalKeyboard' classes. These classes were already challenging due to their small amount of samples and similar harmonic and tonal content, making them prone to confusion. While the data refinement reduced intraclass confusion, it worsened this specific interclass accuracy. To mitigate this phenomenon, the training dataset was supplemented with targeted samples from AudioSet [12]. Table 1 summarizes the distribution of the training samples for each source class after the refinement and augmentation processes.

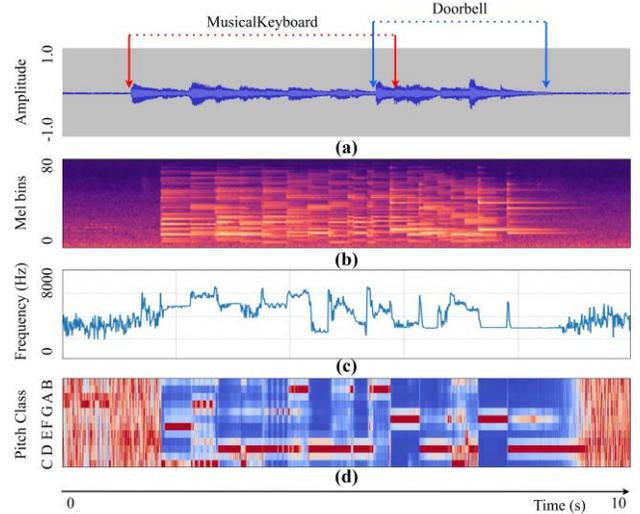

Figure 1: Visualization of audio features used in the proposed system applied to (a) an audio mixture with 'Doorbell' and 'MusicalKeyboard'; (b) mel-spectrogram, (c) spectral roll-off, and (d) chroma features.

## 3. PROPOSED METHOD

In this work, the proposed model is based on the baseline model [10] of DCASE 2025 Challenge Task 4, Masked modeling duo audio-tagging (M2D-AT) and ResUNetK for audio-tagging and source separation, respectively. Our approach attempts to enhance audio-tagging performance by incorporating various audio features and applying an agent-based label correction system to reduce class confusion and FP predictions, respectively.

### 3.1. Spectral roll-off

First of all, we utilize the spectral roll-off feature as an auxiliary input to enhance the ability of the model to distinguish between sound events. The spectral roll-off feature represents the frequency below which a certain percentage of the total spectral energy is contained [4]. This feature captures the distribution of energy across frequencies, providing information about the sharpness or brightness of a sound. For example, impulsive classes, such as 'Clapping' or 'Percussion', tend to have a high roll-off point due to strong high-frequency components, whereas background noises often display lower roll-off characteristics.

By incorporating spectral roll-off, this work aims to provide the model with additional cues for detecting high-frequency transients and differentiating them from continuous or low-frequency sounds. The roll-off feature is extracted framewise and processed via multiple linear layers. The feature is concatenated with the M2D embedding feature [13], resulting in a richer input representation for training the audio-tagging model.

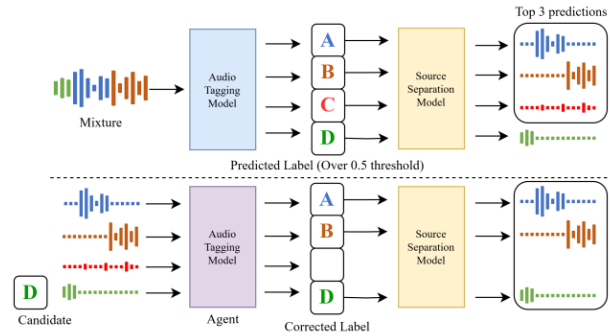

Figure 2: Illustration of agent-based label correction scenario, where {A, B, D} is a set of ground truth labels.

### 3.2. Chroma feature

Second, we try to further enrich the input representation by concatenating the framewise chroma feature with the M2D embedding feature. The chroma feature represents the energy distribution across the 12 pitch classes of the musical octave, making the classes invariant to the absolute frequency [5].

Thus, the proposed audio-tagging model can capture the harmonic and tonal characteristics of audio signals with the help of the chroma feature. In other words, tonal events (e.g., 'AlarmClock') can be separable from atonal noise (e.g., 'HairDryer'). Furthermore, it is expected that spectrally similar tonal sounds, such as 'Doorbell' and 'MusicalKeyboard', should be differentiable from each other due their different harmonic characteristics that are modeled by the chroma feature.

Fig. 1 illustrates each feature applied to a mixture wave (Fig. 1(a)) that is excerpted from a training dataset, composed of 'Doorbell' and 'MusicalKeyboard,' As shown in Fig. 1(a), it is difficult to differentiate two events from mel-spectrogram. However, as shown in Fig. 1(c), the spectral roll-off for 'Doorbell' has higher frequency than that of 'MusicalKeyboard', which enables us to differentiate them. Moreover, the two events have different pitch bands, as shown in Fig. 1(d).

### 3.3. Agent-based label correction

Finally, we propose an agent-based label correction system that operates as a post-processing step after source separation, refining the output of the primary tagging model. The proposed label correction agent takes an estimated source audio and its corresponding audio-tag (Label-1) as input to assess consistency. The estimated source audio is then classified again using the audio-tagging model to get Label-2. If the Label-2 is different from the Lable-1, this tag is removed for the CA-SDRi computation.

Although this correction step is crucial for improving the CA-SDRi metric by reducing FPs, it can also cause an undesirable decrease in the true-positive rate (recall). To this end, the audio-tagging model is modified so that it provides more than three events if their sigmoid scores are higher than a pre-defined threshold, as shown in the top arm of Fig. 2, where {A, B, C} is a set of top-3 predicted labels from audio-tagging model and {D} is additional labels above the threshold. Then, each of them is classified using the audio-tagging model, and re-ranked according to the sigmoid scores, as shown in the bottom arm of Fig. 2, where {A, B,

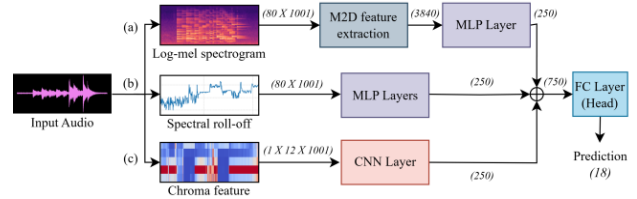

Figure 3: Block diagram of proposed model architectures, where the baseline system corresponds to (a), the spectral roll-off system is constructed using (a) and (b), and the chroma system is constructed using (a) and (c). Note that the spectral roll-off with the chroma system is constructed using all the processing blocks.

D} is re-ranked from {A, B, C}. Finally, three events are taken and then the source separation is carried out again using the new set of event labels.

### 3.4. Model selection metric

#### 3.4.1. Uncorrelated metric challenges

The official evaluation criteria for this task present a challenge for model selection. The audio-tagging model is evaluated on the set-based tagging accuracy, whereas the separation model is evaluated on CA-SDRi. The experiments revealed a weak correlation between these two metrics, making it difficult to determine which tagging model would produce the best separation results. Thus, two supplementary accuracy metrics are designed in this work to create a more holistic evaluation framework.

#### 3.4.2. Macro-averaged accuracy

The limitation of the conventional set-based accuracy is its inability to account for partially correct predictions. To address this problem, we employ a macro-averaged accuracy, which calculates accuracy on a per-label basis rather than a per-set basis. For example, if a model correctly identifies two out of three ground-truth labels, we set its score as 66.7%, while the set-based accuracy is 0%. Consequently, the macro-average accuracy can offer a more granular measure of performance than the set-based accuracy.

#### 3.4.3. False-positive penalized accuracy

The CA-SDRi metric is extremely sensitive to FPs from the audio-tagging model. Thus, we develop an FP-penalized accuracy by incorporating a penalty for FPs directly into the accuracy score to better address this problem. This metric is defined as the number of correctly predicted labels divided by the total number of unique labels in the union of the prediction and ground-truth sets, such as

$$FP\ Panelized\ Accuracy = \frac{TP}{TP+FN+FP}. \quad (1)$$

For instance, given a ground truth of {A, B, C} and a prediction of {A, D, E}, the score is $\frac{1}{1+2+2} = 20\%$. This metric allows the selection of models for audio-tagging to be accurate and less prone to generating FPs.

Table 2: Performance comparison of spatial semantic sound source separation models and the audio tagging model evaluated using standard (Acc1), macro-averaged (Acc2), and false-positive-penalized (Acc3) accuracy metrics.

| Models | | Train DB | w/o agent | | | | with Agent | | | |
|---|---|---|---|---|---|---|---|---|---|---|
| Audio-tagging model | Source-separation model | Refined | Acc1 | Acc2 | Acc3 | CA-SDRi | Acc1 | Acc2 | Acc3 | CA-SDRi |
| Baseline ckpt* | Baseline ckpt* | No | 59.80 | 82.07 | 55.77 | 11.088 | 61.47 | 81.20 | 55.43 | 11.244 |
| Baseline_retrained | Baseline ckpt* | Only AT | 68.80 | 84.50 | 58.41 | 11.380 | 68.53 | 83.33 | 57.64 | 11.400 |
| Baseline_retrained | ResUNetK | Yes | 68.80 | 84.50 | 58.41 | 12.306 | 69.00 | 83.97 | 58.06 | 12.340 |
| + Spectral roll-off | ResUNetK | Yes | 69.13 | 86.20 | 59.69 | 12.328 | 69.80 | 85.80 | 59.44 | 12.369 |
| + Chroma | ResUNetK | Yes | 68.27 | 86.23 | 59.65 | 12.426 | 68.93 | 85.90 | 59.49 | 12.475 |
| + Spectral roll-off + Chroma | ResUNetK | Yes | 69.53 | 86.23 | 59.73 | 12.532 | 69.20 | 85.70 | 59.29 | 12.541 |
| Ensemble** | ResUNetK | Yes | 72.47 | 87.07 | 60.62 | 12.721 | 72.13 | 86.47 | 60.14 | **12.726** |

* Baseline ckpt was released from the organizers of DCASE 2025 Task 4 [14].
** Ensemble model uses *baseline retrained*, s*pectral roll-off*, *chroma*, and *spectral roll-off + chroma*.

## 4. EXPERIMENTAL RESULTS

### 4.1. Model training

The audio-tagging and source-separation models were trained independently. All experiments were conducted on a single Nvidia RTX A6000 graphics processing unit.

#### 4.1.1. Audio-tagging module

The audio-tagging model paired a pretrained M2D [13] backbone with a custom, dual-path classification head that processed the main M2D embedding in parallel with an additional auxiliary audio feature. The architecture of the auxiliary path was conditional on the dimensionality of the feature, as depicted in Fig. 3.

First of all, the spectral roll-off or M2D embedding was processed by two multilayer perceptrons (MLPs) to produce 256-dimensional (256D) embedding vector, as shown in Figs. 3(a) and 3(b). Conversely, the chroma feature was processed by a convolutional neural network consisting of two convolutional layers, resulting in a 256D embedding, as shown in Fig. 3(c). In all configurations, the outputs of the two parallel paths were concatenated and input into a final linear classifier to produce the 18D class logits.

In this work, we adopted the two-stage training strategy used for the baseline to ensure stable learning and effective fine-tuning. In other words, the first stage of the M2D backbone was frozen, and the classification head was trained for 140 epochs with a learning rate of $1\times10^{-3}$ and a batch size of 32. Resuming from the Stage 1 checkpoint, Stage 2 unfroze the last two layers of the M2D backbone and performed joint fine-tuning with the head for 1,500 epochs with a batch size of 128.

The final tagging predictions were generated by a weighted ensemble of the four main models. The weights were assigned based on the validation set performance. The spectral roll-off and chroma features together (0.35) had the most significant influence on the model, followed by chroma feature alone (0.3), spectral roll-off alone (0.2), and the retrained baseline, which was assigned the lowest weight (0.15).

#### 4.1.2. Source-separation module

For the separation task, we employed the ResUNetK checkpoint model from the DCASE 2025 Challenge Task 4 official GitHub [10]. The model was fine-tuned using the refined training dataset.

### 4.2. Discussion

The performance of the proposed audio-tagging model was evaluated using the metrics mentioned in Section 3.4. Table 2 summarizes the performance of different systems, demonstrating that the submitted systems consistently and progressively outperformed the official.

The initial experiments focused on enriching the input representation beyond the standard mel-spectrogram. Including the spectral roll-off feature yielded a modest but consistent improvement across all metrics. Substantial improvement was achieved by adding the chroma feature, resulting in CA-SDRi increment by 0.12 dB. This result confirms that providing explicit audio feature information is beneficial for this task, likely helping the model to distinguish between spectrally similar but harmonically distinct events, such as 'Doorbell' and 'MusicalKeyboard'.

One of the most significant improvements came from the dataset refinement. Auditing the source data to remove ambiguous samples and augmenting specific classes relatively increased CA-SDRi by 10.9% compared to the official baseline. This result highlights a crucial aspect of working with synthesized datasets: model performance is highly sensitive to the quality of the source pool. Improving intraclass consistency and ensuring sufficient samples for all classes provides a much stronger foundation for the model to learn, demonstrated by the tagging and separation performance.

The agent-based correction system presented positive results. Despite the agent occasionally removing true-positive labels, it improved the CA-SDRi by 0.2 dB, demonstrating its effectiveness in reducing the FPs that were heavily penalized by this metric. This trade-off between recall and precision highlights that optimizing purely for tagging accuracy does not guarantee the best performance on the S5 task. A custom metric, the FP-penalized accuracy, was developed to navigate this balance better during model selection, enabling a more effective evaluation of models based on their suitability for the CA-SDRi-driven goal. The overall positive influence of the agent on the primary competition metric justifies its inclusion as a valuable, task-aware post-processing step.

## 5. CONCLUSION

This technical report presented the submission systems for the DCASE 2025 Challenge Task 4, focusing on a multifaceted approach to improve spatial semantic segmentation. The proposed strategy combined additional audio feature input (spectral roll-off and chroma), a dataset refinement process, and an agent-based error correction system. The experimental results demonstrated the value of this holistic approach. Each component contributed incrementally to the final performance, culminating in a system that significantly outperformed the baseline with a 14.7% relative increase in CA-SDRi. Moreover, careful data refinement yielded substantial gains, and task-aware post-processing was crucial for optimizing the final competition metric, even if they resulted in a slight trade-off with intermediate tagging accuracy.